\documentclass{article}

\usepackage{amsmath,amsthm}
\normalsize \rm
\begin{document}
\begin{center}
{\huge \textbf {A sensible proof connecting the scale-free feature with the Zipf-law}}\\[12pt]
{\large Fei Ma$^{a,}$\footnote{~The author's E-mail: mafei123987@163.com. }}\\[6pt]
{\footnotesize $^{a}$ School of Electronics Engineering and Computer Science, Peking University, Beijing 100871, China\\[12pt]}
\end{center}

\begin{quote}
\textbf{Abstract:} Most of various large-size complex systems in nature and society can be well described as complex networks (graphs) to better understand the evolutional mechanisms and dynamical functions behind themselves. Of some part follow scale-free behavior, that is, the ratio of the number of vertices with degree more than or equal to $k$ and order of the whole network obeys the expression $P_{cum}(k)\sim k^{1-\gamma}$ ($2<\gamma<3$). Meanwhile, the Zipf-law, which satisfies this $f_{r}\sim r^{-\alpha}$ ($\alpha$ close to unity), is also prevalent in many complex systems, such as word frequencies in text and city sizes. It can be easily noticed that the both above have same type of appearance, namely the known power-law. Compared to the scale-free feature proofed analytically by continuum theory, by far the latter in most cases still is thought of as an empirical principle in lots of science communities, particularly in social science. For this reason there is a need for either pointing out the inner connection between the two or distinguishing difference of one another. Here, for any arbitrary given scale-free network model of order $N$, we report an equivalent relation between scale-free feature and the Zipf-law based on the vertex rank. By rigorous mathematical derivations, we eliminate the gap, lack of theoretical fundament of the Zipf-law. Therefore one can be convinced that it is reasonable to adopt methods already used to study complex networks to do the Zipf-law .


\end{quote}

\vskip 1cm

\section{The first proof}

In most real-life instances, each vertex in complex network with the same degree value plays different role and implements various function. Such an example is word's rank in the long-scale text. In this situation, each word is assigned a unique number based on its own frequency. George Kingsley Zipf \cite{G-K-Z-1949} addressed that the frequency $f_{r}$ with which words are used seems to meet a power law

$$f_{r}\sim \frac{C}{r^{\alpha}}$$
in which symbol $C$ is an invariant and $\alpha$ is approximately equal to unity.

 With this rank hypothesis, the Zipf-law is popular enough around our daily-life world, such as word rank, etc. However there are no known proofs to clearly answer the reason for emergence of the Zipf-law such that it is still thought of as an empirical principle formulated using mathematical statistic. Although many published literatures attempt to seek for reasonable explorations, a rigorous and accessible mathematical proof can be not acquired. It is indeed a challenging and demanded work to bring a precise solution to identify the Zipf-law. Fortunately, in the following, we put forward a compacted proof. To do this, we have to recall the concept of scale-free model \cite{Albert-1999-1}.

For a given scale-free model of order $N$, we immediately see

\begin{equation}\label{eqa:3-3-1}
P(k)\sim k^{-\gamma},  \qquad N_{k_{i}}=NP(k_{i})=\frac{N}{k_{i}^{\gamma}}.
\end{equation}

Based on unique rank number $r$ corresponding vertex frequency $f_{r}$, we can randomly select a vertex in rank list composed of all vertices. Because of both commonly seen phenomena, rank number being a continuous natural number sequence from 1 to the maximum value $N$ and yet the degree (frequency) sequence not to be so, possible for a continuous integer interval to consist of many vertices with same degree (frequency). Without loss of generality, list all vertex in decrease order of vertex-degree. When choose a vertex $i$ of degree $k_{i}$ at random, its rank number $r_{k_{i}}$ must fall into this range between $\sum_{k_{i}+\delta^{i}_{+}}^{k_{max}} N_{k_{i}}$ and $\sum_{k_{i}+\delta^{i}_{-}}^{k_{max}} N_{k_{i}}$, namely

\begin{equation}\label{eqa:3-3-2}
\sum_{k_{i}+\delta^{i}_{+}}^{k_{max}} N_{k_{i}}\leq r_{k_{i}}\leq \sum_{k_{i}-\delta^{i}_{-}}^{k_{max}} N_{k_{i}}
\end{equation}
where symbol $N_{k_{i}}$ denotes the number of vertices of degree $k_{i}$, $k_{max}$ the maximum degree value, $\delta^{i}_{+}$ the difference equal to the minimal degree value, which belongs to $k_{i}$'s left neighbor set in degree sequence and is more than $k_{i}$, subtracting $k_{i}$, as well $\delta^{i}_{-}$ the difference equal to $k_{i}$ subtracting the maximal degree value , which is in $k_{i}$'s left neighbor set in degree sequence and less than $k_{i}$. If consider all degree value as continuous variables, combining Eq.\ref{eqa:3-3-1} and Eq.\ref{eqa:3-3-2} yields

\begin{equation}\label{eqa:3-3-3}
N\int_{k_{i}+\delta^{i}_{+}}^{k_{max}}k^{-\gamma}dk\leq r_{k_{i}}\leq N\int_{k_{i}-\delta^{i}_{-}}^{k_{max}}k^{-\gamma}dk.
\end{equation}

Using elementary integral calculations, we have

\begin{equation}\label{eqa:3-3-4}
\frac{N}{1-\gamma}(k_{max}^{-\gamma+1}-(k_{i}+\delta^{i}_{+})^{-\gamma+1})\leq r_{k_{i}}\leq\frac{N}{1-\gamma}(k_{max}^{-\gamma+1}-(k_{i}-\delta^{i}_{-})^{-\gamma+1}).
\end{equation}

Taking into consideration in any large-scale network the maximal degree value $k_{max}$ being several orders of magnitude in comparison with other degree values, we obtain asymptotically the following inequality by omitting the influence from $k_{max}$ at both sides of Eq.\ref{eqa:3-3-4}, as follows

\begin{equation}\label{eqa:3-3-5}
\frac{N}{\gamma-1}(k_{i}+\delta^{i}_{+})^{-\gamma+1}\preceq r_{k_{i}}\preceq \frac{N}{\gamma-1}(k_{i}-\delta^{i}_{-})^{-\gamma+1}.
\end{equation}

Generally, the smaller both $\delta^{i}_{+}$ and $\delta^{i}_{-}$ are, the closer to $|V|$ rank seats are. In degree value density regions, either $\delta^{i}_{+}$ or $\delta^{i}_{-}$ can visit at the minimal value $1$. Hence, it is available that we may keep approximation, having

\begin{equation}\label{eqa:3-3-6}
k_{i}-\delta^{i}_{-}\preceq \left(\frac{N}{\gamma-1}\right)^{\frac{1}{\gamma-1}}r_{k_{i}}^{\frac{1}{1-\gamma}}\preceq k_{i}+\delta^{i}_{+}.
\end{equation}

Again making further approximation yields

\begin{equation}\label{eqa:3-3-7}
k_{i}\preceq \left(\frac{N}{\gamma-1}\right)^{\frac{1}{\gamma-1}}r_{k_{i}}^{\frac{1}{1-\gamma}}\preceq k_{i}.
\end{equation}

Thus, we obtain

\begin{equation}\label{eqa:3-3-7}
f_{r_{k_{i}}}=k_{i}\sim \frac{C}{r_{k_{i}}^{\alpha}}.
\end{equation}
here $C=\left(\frac{N}{\gamma-1}\right)^{\frac{1}{\gamma-1}}$ and $\alpha=\frac{1}{\gamma-1}$. This is complete.

With our initial assumption for parameter $\gamma$, $2< \gamma< 3$, the value of index $\alpha$ will naturally fall into the region $\frac{1}{2}< \gamma< 1$. The closer to 2 the degree exponent $\gamma$ is, the closer to unity the frequency index $\alpha$ is, showing directly which our result is in considerable agreement with the description of the Zipf-law. Not only so, we also provide a measure to asymptotically compute a value for parameter $C$ of the Zipf-law. To close our here discussions and highlight our main work, the relations among Pareto distribution $P_{cum}(x)$, scale-free feature $P(x)$ and Zipf-law $f_{r}$ should be illustrated, as follows

\begin{center}
\vspace*{-1mm}$$ P_{cum}(k)\sim k^{1-\gamma}\stackrel{first-order \:integral}\longleftarrow degree\:distribution\: P(k)\sim k^{-\gamma}\stackrel {first-order \:differential }\longleftarrow  P_{cum}(k)\sim k^{1-\gamma}.    \vspace*{0mm}$$
\end{center}

\vspace*{-2mm}$$ The\: Zipf-law\: f_{r}\sim \frac{C}{r^{\alpha}}\stackrel{exchanging}\Longleftrightarrow  degree\:distribution\: P(k)\sim k^{-\gamma}.      \vspace*{-2mm}$$

\section{The second proof}

In a scale-free network $G=(V,E)$ \cite{Albert-1999-1}, degree distribution $P(k)$ follows 

$$P(k)\sim k^{-\gamma},$$
where power-law exponent $\gamma$ is no less than $1$. The network is considered sparse if exponent $\gamma$ is strictly more than $2$. This is Power-law in the filed of complex networks. 

It is widely known that given a corpus \cite{G-K-Z-1949}, the frequency $f_{v}$ of verb $v$ and its rank $r_{v}$ asymptotically obey 

$$f_{v}\sim  r_{v}^{-1}.$$
This is famous Zipf-law due to G.K. Zipf. 

\emph{\textbf {Theorem } Power-law is equivalent to Zipf-law.} 

To prove the correctness of the theorem above, we need to introduce two lemmas as below.

\emph{\textbf{Lemma 1} Given a sparse scale-free network $G=(V,E)$, the ratio $\alpha(k)$ of the summation of degree of vertices whose degree is no less than $k$, denoted by $A_{k}$, to the number of edges $|E|$ is given by} 

\begin{equation}\label{eqa:1}
\alpha(k)=\frac{A_{k}}{|E|}=O\left(k^{-\lambda}\right),\qquad  \lambda=\gamma-2.
\end{equation}   

\emph{Proof} By definition, quantity $A_{k}$ is given by

\begin{equation}\label{eqa:1-1}
A_{k}=\sum_{j\geq k}n_{j}=|V|\sum_{j\geq k}kP(k),
\end{equation}  
in which $n_{j}$ presents the summation of degree of vertices whose degree is exactly equal to $j$. Then, we have 

\begin{equation}\label{eqa:1-2}
A_{k}\sim \frac{|V|}{\gamma-2}k^{-\gamma+2}.
\end{equation}
Note that we have used 
$$k_{max}=O\left(|V|^{\frac{1}{\gamma-1}}\right).$$

Next, we come to 

\begin{equation}\label{eqa:1-2}
\alpha(k)\sim\frac{|V|}{\gamma-1}k^{-\gamma+2}/O\left(|V|\right)=O\left(k^{-\lambda}\right).
\end{equation}
Notice also that we have used the following fact
$$|E|=O\left(|V|\right).$$
This is complete. \qed 

\emph{\textbf{Lemma 2} Given a sparse scale-free network $G=(V,E)$, the ratio $\beta(k)$ of the summation of degree of vertices whose degree is no less than $k$, denoted by $A_{k}$, to the number of vertices of this kind $|V_{k}|$ is given by} 

\begin{equation}\label{eqa:2}
\beta(k)=\frac{A_{k}}{|V_{k}|}=O\left(k^{-\xi}\right),\qquad  \xi=1.
\end{equation} 

\emph{Proof} It is easy to prove Lemma 2 based on Lemma 1 and the definition of $|V_{k}|$. This is complete. \qed  

From Lemma 1 and 2, it is clear to see that Theorem holds true.

\section{Discussion and conclusion}
 Our results in some extent are a significant expansion of previously excellent achievements, meanwhile can be viewed as a perfect theoretical integration. Based on rigorous mathematical derivation, one should be convinced that now the scale-free feature and the Zipf-law communicate with one another in complex systems. Although here reports our recent work, it is just a tip of the iceberg. We always firmly believe that there will be still more challenges and difficulties to be overcome before better understanding the considerable potential from the power-law both experimentally and theoretically.

\end{document}